\documentclass[prb,preprint]{revtex4-1} 

\usepackage{amsmath}  
\usepackage{amsfonts} 
\usepackage{graphicx} 
\usepackage{color}
\usepackage{ulem}

\def\source#1#2#3#4{{\it #1}~{\bf #2}, #3 (#4)}
\def\be{\begin{equation}}
\def\ee{\end{equation}}
\def\Eq#1{Eq. \ref{#1}}

\def\ie{i.e.}

\def\ket#1{| #1 \rangle}

\begin{document}


\title{The original Wigner's-Friend scenarios}

\author{Jay Lawrence}
	\email{jay.lawrence@dartmouth.edu}
	 \affiliation{Department of Physics and Astronomy, Dartmouth College, 
	 Hanover, NH, USA} 
\altaffiliation[permanent address: ]{P O Box 965, Grantham, NH, USA} 


\date{\today}

\begin{abstract}
We describe the Wigner's-Friend scenario according to Wigner, then a 
similar but earlier version according to Everett.
Wigner and Everett prcvide different resolutions of
essentially the same paradox.  Decoherence theory provides a third
resolution.  Despite different interpretations (or their absence), these 
three stories fit together to form a consistent picture without a paradox.

\end{abstract}

\maketitle 

\section{Introduction} 
\noindent  There has been a surge of interest over the past decade in the 
subject of ``Wigner's Friend,'' referring to a brief scenario presented by 
E. P. Wigner in 1961.\cite{Wigner} This accompanies a broader expansion 
of interest in foundational issues, which includes the emergence of new 
observer-centric interpretations of quantum theory, such as the 
relational interpretation\cite{Rovelli} and the QBist approach.\cite{QBist} 
These developments are encouraged by the emergence and recognition 
of the new subfield, Quantum Information, which spans foundational 
issues along with the more practical pursuits of quantum computation 
and quantum communication.\cite{APS status}  

The work on Wigner's-Friend-like scenarios\cite{Schmid.23}  in particular 
addresses a subset of issues within the larger measurement problem. 
These include the roles of observers in measurement, the consistency 
of information available to multiple observers, and the 
objectivity of the wavefunction and of quantum mechanics itself.

In this paper I will focus on the two original scenarios - Wigner's scenario 
of 1961, and an earlier (lesser known) scenario by Hugh Everett in 1956.
\cite{sketches,Barrett} Both of them were motivated by a paradox (then
unnamed) whose purpose was to expose contradictions lurking in the 
naive application of quantum theory to systems with multiple observers. 
But the real messages were in the proposed resolutions.  While the  two 
stated paradoxes were essentially the same, the resolutions were very 
different.  The more specific aims of this paper are to articulate these 
resolutions in the context of the authors' beliefs at the time, and 
identify the foundational issues that are still in play in current 
Wigner's-Friend-like investigations. It is particularly important to 
cite Everett's contribution, as I suspect that many readers will not 
be aware of {\it his} Wigner's-Friend story, which only appeared in a 
preliminary version of his Ph.D. thesis.\cite{sketches}

I will show how the two very different resolutions fit together
coherently, and add a supporting third resolution based on the theory of
decoherence.  The last provides a more current perspective while adding 
important content.  I make no attempt to review the current literature.  
Our intent here is to provide a conceptual background, augmented 
by Everett's original contribution, for those readers who wish to 
further explore the literature, or simply to attend colloquia based on 
current work.  It could also be useful to teachers in motivating 
discussions of foundations and interpretations of quantum mechanics.

The three resolutions will be presented in Sections II, III, and IV. 
To set the stage, here is a bare-bones sketch that defines the paradox, 
but stops short of the resolutions.  The first observer (Wigner's Friend 
$F$) performs a measurement on the system $S$, which has been 
prepared in a superposition of its possible output states - \ie, 
eigenstates of the measured observable. Assuming for convenience 
that $S$ is a two-state system, the prepared state is
\be
   \ket{\psi}_s = \alpha \ket{0}_s + \beta \ket{1}_s.
\label{prepared}
\ee
The Friend $F$ obtains a definite answer; he finds the state to be
either $\ket{0}_s$ or $\ket{1}_s$, not a superposition.  A second 
observer, Wigner ($W$), does not observe $F's$ measurement 
directly, but, applying quantum theory to the joint system 
$SF \equiv S \otimes F$ and assuming the linearity of quantum 
dynamics for the (unobserved) system, he concludes that the state 
of  $SF$ must be the superposition
\be
   \ket{\Psi} = \alpha \ket{0}_s \ket{0}_f + \beta \ket{1}_s \ket{1}_f,
\label{equationSF}
\ee
where $\ket{0}_f$ is the state in which $F$ perceives the outcome 
to be 0, and $\ket{1}_f$ is the alternative, the perception of 
outcome 1.  The difference in state assignments - a definite state 
seen by $F$ ($\ket{0}_s$ or $\ket{1}_s$), versus the superposition 
(\Eq{equationSF}) inferred by Wigner - is called the 
Wigner's-Friend Paradox.

We now turn to a discussion of the individual cases, including 
nuances in the paradox, followed by the resolutions, which are 
radically different. 

\section{Resolution according to Wigner}

\noindent E. P. Wigner presented his scenario in the article, 
``Remarks on the Mind-Body Question,'' which appeared in the 
volume {\it The Scientist Speculates},\cite{Wigner} by I. J. Good 
(1961).  The main theme of this article is the role of consciousness 
in quantum mechanics.  Wigner believed that consciousness is 
the system uniquely capable of collapsing the wavefunction:  The 
observer gains an ``impression'' of the value of the measured 
quantity, and when this impression enters the consciousness, the 
wavefunction collapses.\cite{commentary}  
In Wigner's scenario, the Friend's measurement consists of 
looking in a prescribed direction, at a prescribed time, and either 
seeing - or not seeing - a flash. Following this measurement, 
Wigner asks $F$ if he saw a flash:  The answer is definite - a 
definite ``Yes!'' or a definite ``No!'' as the case may be.  Wigner
then asks $F$ whether he was confused before answering the 
question:  The answer is a definite ``No!''  The friend knew the 
result before the question was asked.  From this, Wigner 
concludes that $F$, and not Wigner himself, collapsed 
the wavefunction.

Wigner admits that this collapse is a ``highly non-linear'' event.  
For Wigner, it marks a violation of linearity by the intervention of
$F$'s consciousness, and a case of the mind directly affecting 
matter.  He points out that, mathematically, the initial superposition 
is converted to a mixture, which is represented by the density 
matrix, not a wavefunction.  Accordingly, $F$ sees a definite 
outcome, with a probability given by the Born rule.  

 
Physicists have not generally accepted Wigner's idea on the active role 
of consciousness.  What is relevant here, however, is not the specific
mechanism of collapse, but rather Wigner's recognition that collapse 
occurs without his agency.   This resolves the Wigner's-Friend Paradox, 
if not the larger measurement problem.  And it is consistent with the 
common (if not universal) understanding of the collapse axiom in 
standard quantum mechanics,\cite{standard,vN.55,Sakurai,Weinbergbook} 
which holds that a collapse occurs upon any measurement, 
by any observer.  For readers unsatisfied by the collapse axiom, 
the next two examples (Secs. III and IV) are interesting.


\section{Resolution according to Everett}

\noindent  It is not widely known that Hugh Everett actually authored the 
first Wigner's-Friend scenario - he wrote this out in a preliminary version 
(the so-called ``long version'') of his PhD thesis at 
Princeton\cite{sketches,Barrett} in 1956, a half-decade prior to Wigner's 
publication.  His purpose was to illustrate, and to resolve, contradictions 
which arise when two observers appear in the wavefunction (he calls 
them $A$ and $B$ - nowadays we would call them Alice and Bob). 

Observer $A$ conducts his measurement in a 
closed, isolated room,\cite{isolated}
using a measuring apparatus, and he records his result in a notebook.
Observer $B$, isolated from the measurement, is in possession of the 
initial state of the room and its contents prior to the measurement, and 
he uses Schr\"{o}dinger's equation (\ie, unitary evolution) to evaluate 
the post-measurement state, obtaining \Eq{equationSF} for $SA$.  This 
version of the contradiction, with its greater attention to detail, is closer 
to the current conception of the paradox.

Much later, $B$ enters the room and reads the entry in $A$'s notebook.   
Unlike Wigner, $B$ believes that he alone, being ``external,'' collapsed 
the wavefunction of $SA$.  He expects $A$'s gratitude for this action, 
which (in his view) resolved the ambiguity in the state of $SA$.  But 
instead, $A$ rudely rejects $B$'s opinion, asserting that he knew the 
result from the beginning.   He further punctures $B$'s ego by pointing 
out that if $B$ were right, then a third observer could similarly render 
the current situation an illusion.  Here, Everett rests his case.

Now for Everett's resolution: He abandons the collapse axiom.  He 
places all observers in the wavefunction on equal footing with other
quantum objects, giving them no privileged position as ``outsiders.''
He models an observer as a machine with a memory, which can
read and record the output of an apparatus.  He uses a single index 
(as in \Eq{equationSF}) for both the apparatus state and the 
observer's memory state.  Here, to emphasize that apparatus and 
observer are separate systems, we label both states (redundantly);
\be
   \ket{\Psi} = \alpha \ket{0}_s \ket{0,0}_a + 
   \beta \ket{1}_s \ket{1,1}_a,
\label{equationSA}
\ee
where the first index in $\ket{k,k}_a$ is the reading of the apparatus, 
and the second is the observer's memory state,\cite{many} showing 
that $A$ read the apparatus correctly.

Everett's crucial argument is that the perfect correlations expressed 
by (\ref{equationSA}) imply that within each term,  $A$ is 
aware of only the one reading represented therein; he is unaware 
of the existence of an alternate term with a different reading.  Thus 
he {\it experiences} a collapse, even though there is no actual 
collapse of the $SA$ wavefunction.  His observed collapse, and 
the associated Born rule probability, are subjective. The lack 
of observable interference between terms is referred to 
as a {\it branching}\cite{branching,Taylor} 
of the wavefunction, and, as in the above case, it is properly 
expressed in terms of a density matrix, not a wavefunction.



To see the relationship between the Everett and Wigner pictures, 
note that Everett enlarged the memory states of observers to 
include multiple measurements in sequence - either a sequence 
of measurements on identically prepared systems, or a repeated
measurement on the same system.  From such considerations, 
he derived the familiar probability interpretation rules of standard 
quantum mechanics, which could be inferred by his model 
observer.  So an actual human observer, who (like the model) 
can perceive only a single outcome, would experience the world 
according to the {\it standard} formulation, consistent with 
Wigner's picture.

Everett's work culminated in his Relative State Interpretation 
of quantum mechanics - the subject of his 1957 PhD thesis
(the ``short version''),\cite{thesis}  and his seminal 
paper\cite{Everett} of the same year.  His Wigner's-Friend story 
does not reappear in either of these later works.  We do not 
wish to attempt here a comprehensive account of Everett's 
formulation or the more developed Many Worlds
Interpretation,\cite{deWitt} but we do recommend Everett's 
earlier version\cite{sketches} and the commentary included 
in the evolume by Barrett and Byrne.\cite{Barrett} 

\section{Resolution in Decoherence}

\noindent The term ``decoherence theory,'' refers to a general  approach 
to the loss of coherence, due to interactions with the environment, 
between terms in a superposition state of a system.  It is marked by the 
system's inability to display superpositions.  The theory employs 
textbook quantum mechanics,\cite{Sakurai,Weinbergbook} except 
that, like Everett, it abandons the collapse axiom because it aims to
derive such effects through (unitary) Schr\"{o}dinger evolution. This
theory is, however, {\it unlike} Everett's in its attempt to account for 
environmental effects on the system of interest (be it microscopic or 
macroscopic), and it's removal of observers from the wavefunction.  

In this paper, we refer specifically to a theory\cite{Zeh,Zurek.81} of 
the environmental effect on the {\it measuring apparatus}.  For the 
apparatus to function properly, decoherence must suppress the 
interference between different pointer states.  So this theory 
shifts the focus away from the observer to the apparatus and its 
environment.  As elaborated in the paragraphs below, there is an
apparent collapse of the wavefunction, now manifested in the 
definite position taken by the pointer.  Thus, the transition from a
wavefunction to a density matrix occurs before the observer 
enters the picture.  The observer's role here is to read the pointer
state displayed by the apparatus.  His perceptions and memories 
of these states, here considered classically, are the same as 
they would be in the Everett model.

The above results were established, conceptually and technically, by 
two seminal works.  First, in 1970, H. D. Zeh \cite{Zeh} emphasized 
the need to account for the macroscopic nature of the apparatus.  
He argued that such a system cannot exist in a pure state.  It must
be described by a density matrix - one that is diagonal in the output
states of the system.  This forms a mathematical statement of the 
axiom of measurement - it expresses what we see - a definite but 
random output chosen from among the allowed ``pointer'' states 
of the apparatus.  These states are analogous to the live 
and dead states of Schr\"{o}dinger's Cat.  

Then, in 1981, W. H. Zurek\cite{Zurek.81} argued that the basis of 
possible output states 
is determined by the apparatus-environment interaction, with the
understanding that the environment may consist of internal degrees 
of freedom of the apparatus as well as the external environment.  In a
``tour-de-force''  model calculation of the joint apparatus-environment 
system,\cite{Zurek.82} he traced out the environmental degrees of 
freedom to obtain the reduced density matrix of the pointer.  His 
calculation showed that this is diagonal in the basis of allowed 
pointer states to an excellent approximation,\cite{offdiagonal} so 
that only a single output is displayed for the observer.

Reflecting on the above results, the dynamical evolution of the 
system-apparatus-environment is unitary, so that the final state 
is a superposition of all possible outputs. But interference 
between them is undetectable, because each pointer state is
accompanied by a different environmental state representing an
enormous number of degrees if freedom. This situation is correctly
described, as in the previous cases, by a density matrix.  In this
case, Zurek has made the transformation explicit with the trace 
operation - the transformation from the wavefunction of the
system-apparatus-environment to the reduced density matrix of 
the system-apparatus.  This transformation is non-unitary, but the 
trace does not represent a physical process; it represents the 
practical limitation of information accessible to the observer - 
the information available on a single element of the 
reduced density matrix.

Because of the survival of all possible outputs in the final wavefunction,
Everett's Interpretation may seem the most natural for decoherence theory, 
but not all practitioners adopt it.  Many choose not to take a position on
the question of the existence of the unobserved branches - a question which 
is currently untestable experimentally.  A few others believe, in contrast, that 
the unobserved branches may be removed by some unknown dynamics.  In 
particular, model non-unitary interactions have been proposed\cite{OCT} that 
do this, although there is no evidence to date for such an interaction. And so
it is, that the question of interpretation remains unsettled.

\section{Epilogue}

\subsection{about Wigner}

Although Wigner's proposal that consciousness causes
collapse is not generally accepted, his more general opinions are
reflected in some current observer-centric interpretations.  The 
idea that consciousness is the primary reality, and that therefore
the wavefunction reflects the content of the consciousness, finds
resonance with current variations on the Copenhagen Interpretation.
\cite{Cabello} Typical of these, the wavefunction represents the
``knowledge of the observer'' (or in QBist doctrine, it represents 
the ``beliefs of the agent''), to be used for the prediction of the 
next ``observations.''  So the wavefunction is not an intrinsic 
property of the system of interest, but rather an instrument for 
making these predictions.

As for Wigner himself, he continued throughout the 1960s and 
early 70s writing about his ideas on consciousness and the 
foundations of quantum theory.  But eventually he changed his 
mind, persuaded by Zeh's paper of 1970\cite{Zeh} and subsequent 
work.\cite{Esfeld}  A convincing aspect was the understanding that
a macroscopic apparatus, treated quantum mechanically with 
account for the environment, could behave as an effectively 
classical object.  Wigner had never been able to accept Bohr's
assertion that an apparatus is {\it intrinsically} classical.\cite{Bohr}

\subsection{about Everett}

Bohr and his colleagues did not accept Everett's theory, despite 
his personal efforts to convince them.\cite{Barrett} Everett's 1957
paper\cite{Everett} was his last publication in Physics.  And yet,
his legacy is far-reaching.  The Many-Worlds Interpretation is
actively pursued, and Google ranks it in the top three of ``preferred 
interpretations,'' along with Copenhagen (first) and hidden variables 
theories. 
The support among cosmologists is strong, with interest in the 
universal wavefunction.

\subsection{about decoherence theory}


Both Zeh and Zurek acknowledge the multiple-branch outcome of 
their decoherence calculations.  Ironically, Zeh was initially unaware 
of Everett's work, and rediscovered his results, generalizing to 
include stability considerations in the determination of pointer 
states,\cite{Camilleri} within his pursuit of decoherence theory.  
Zeh's thinking is relfected in his 1970 paper, where he refers to the
two pointer states arising in a measurement on a two-state system: 
``The different components represent two completely decoupled 
worlds. This decoupling represents exactly the `reduction of the
wavefunction.' As the `other' component cannot be observed
any more, it serves only to save the consistency of quantum 
theory.  Omitting this component is justified pragmatically, but it
leads to the discrepancies discussed above.''\cite{embrace}

Zurek accepts multiple branches as a result of standard
quantum theory without the collapse axiom, but not necessarily 
as the ultimate truth.  He expresses doubts, remarking that many 
questions ``can be answered without having to decide whether, 
where, when or how the ultimate collapse occurs,''\cite{ultimate}  
a nod to the Copenhagen interpretation.

For an excellent accessible discussion of decoherence theory 
and its relationship to Everett and Copenhagen Interpretations,
I recommend Weinberg's recent textbook.\cite{Weinbergpages}
I also recommend Schlosshauer's review article of 2005\cite{Schloss}
on the relevance of decoherence for foundational questions in
general.

\subsection{decoherence versus isolation}

For a more practical comment on decoherence, let us return to the
footnote\cite{isolated} that criticized Everett for his fanciful reference
to a measurement conducted in a ``closed, isolated'' room.  Zeh's 
contention that macroscopic systems cannot be isolated is confirmed
by calculations of Joos and Zeh\cite{Joos} (1985) on decoherence 
effects in a variety of environments, on systems ranging in size from 
large molecules to macroscopic. Even in outer space, the microwave 
background radiation is sufficient to decohere the states of macroscopic
and smaller systems, so that a measuring apparatus would still function.  
It is noteworthy that a measurement itself (as distinct from the 
premeasurement setup) does not require the presence of an observer.  
It is no contradiction that detectors work in outer space.

\section{Conclusions}
The Wigner's-Friend Paradox, as illustrated in the Introduction, is the 
difference in state assignments made by the Friend (an observer who has 
performed a measurement on $S$) and Wigner (a theorist who predicts 
the state of $SF$ from outside).  Wigner\cite{Wigner} resolved the paradox 
by concluding that an observer - one who is part of the system being 
described by quantum theory - is capable of collapsing the wavefunction.  
Everett\cite{sketches} resolved it by abandonding the collapse axiom and 
arguing that, despite the (theoretical) survival of all branches of the 
wavefunction, the Friend perceives only a single outcome. Decoherence 
theory (as originally developed\cite{Zeh,Zurek.81,Zurek.82}) similarly 
rejects the collapse axiom and, focusing on the apparatus rather than the 
observer, demonstrates that environmental decoherence of apparatus 
states causes the apparent collapse, before the observer enters the 
picture.  

The three resolutions above offer three different perspectives on the 
same scenario.  Wigner's perspective is that of the ``real world'' in 
which we live.  We experience the collapse of the wavefunction when 
we perform a measurement.  Everett's perspective is that of the 
universe according to quantum theory without the collapse axiom, 
in which the wavefunction branches upon measurement into 
separate non-interacting ``worlds.''  This is a universe we can only 
imagine.  But from Everett's theory of it, we can derive the collapse 
experienced by real observers (like ourselves) who live in (any) 
one of these real worlds.   The perspective of decoherence theory 
is, like Everett's, that of a universal wavefunction.  But, unlike 
Everett's, among the technical differences mentioned above, it 
takes an agnostic position on the existence of the unobserved 
branches.  From either this {\it or} Everett's perspective, our 
consciousness (like Wigner's in the end) is liberated from 
having to induce the collapse.


\acknowledgements
I would like to thank Miles Blencowe for stimulating discussions of 
Wigner's Friend, Schr\"{o}dinger's Cat, and Decoherence.
\medskip

\noindent DECLARATION: The author has no conflicts to disclose.


\begin{thebibliography}{99}

\bibitem{Wigner} E. P. Wigner, Remarks on the mind-body question, in 
      {\it The Scientist Speculates}, (Ed.) I. J. Good (Heinemann, London, 1061).
\bibitem{Rovelli} C. Rovelli, Relational Quantum Mechanics, 
       \source{Int. J. Theor. Phys.}{35}{1637}{1996}.
\bibitem{QBist} C. A. Fuchs, N. D. Mermin, and R. Schack, An introduction to 
       QBism with an application to the locality of quantum mechanics, 
       \source{Am. J. Phys}{82}{749}{2014}.
\bibitem{APS status} Quantum Information was an APS Topical Group in  
       2005, achieving Divisional Status in 2017.
\bibitem{Schmid.23} D. Schmid, Y. Ying, and M. S. Leifer, A review and analysis 
       of six extended Wigner's friend arguments, arXiv:2308.16220v3[quant-ph].
\bibitem{sketches} H. Everett, The theory of the universal wavefunction, preliminary
version of PhD thesis at Princeton University, reprinted in the ebook by Barrett and 
      Byrne.\cite{Barrett}  Everett's ``Wigner's-Friend'' scenario appears on pp. 73-4 of
      that book.  Also see the related sketch on p. 15.
\bibitem{Barrett} J. A. Barrett and P. Byrne (Eds.),  {\it The Everett Interpretation of 
      Quantum Mechanics: Collected works 1955 - 1980 with commentary.}  
      Princeton University Press (2012).
\bibitem{commentary} Wigner does not mention an apparatus, and his language
      suggests that he may consider the Friend's measurement to be without one.
      One must say that this seems unrealistic; it is difficult to imagine a system 
      described by \Eq{prepared} giving rise to a perceptible flash without 
      amplification by an apparatus.
\bibitem{isolated} The room is {\it closed} so that one can apply the Schr\"{o}dinger 
       equation, and {\it isolated} so that observer $B$ cannot inadvertantly collapse
       the wavefunction.  Neither of these conditions is achievable in practice (see the
       next section).  But if Everett's statement of the paradox is fanciful, his proposed
       resolution is very serious. 
\bibitem{standard} We shall refer to {\it standard} quantum mechanics as the 
      formulation found in most textbooks, sometimes called the Dirac-von Neumann
      formulation,\cite{vN.55} because this includes the collapse axiom among listed
      axioms.  For a particularly good concise account see Sakurai\cite{Sakurai}; for a
      more comprehensive account see Weinberg.\cite{Weinbergbook}  
\bibitem{vN.55} J. von Neumann, {\it Mathematical Foundations of Quantum 
       Mechanics}, trans. R. T. Beyer, Princeton University Press, Princeton (1955).
\bibitem{Sakurai} J. J. Sakurai, {\it Modern Quantum Mechanics}, Addison Wesley, 
      New York (1985).
\bibitem{Weinbergbook} S. Weinberg, {\it Lectures on Quantum Mechanics}, 
      Cambridge University Press (2013).
\bibitem{many} Everett actually considers a many-state system, so that Eqs. 
      \ref{prepared}, \ref{equationSF}, and \ref{equationSA} are all superpositions
      of $n$ terms.  We shall continue to assume just the two terms for simplicity.
\bibitem{branching} Taylor and McCullough\cite{Taylor} define {\it branching} as 
      the inability to distinguish a superposition state from a mixture, as represented 
      by a density matrix.
\bibitem{Taylor} J. K. Taylor and I. P. Mc Culloch, \source{Quantum}{9}{1670}{2025}.
\bibitem{thesis} H. Everett, PhD thesis, Princeton University, Princeton, NJ (1957).
\bibitem{Everett} H. Everett,  ``Relative State'' formulation of quantum mechanics,
      \source{Rev. Mod. Phys.}{29}{454}{1957}.
\bibitem{deWitt} B. S. DeWitt and N. Graham, {\it The Many-Worlds Interpretation 
       of Quantum Mechanics}, (Princeton University Press, Princeton, 1973).
\bibitem{Zeh} H. D. Zeh, On the interpretation of measurement in quantum theory,   
       \source{Foundations of Physics}{1}{69}{1970}.
\bibitem{Zurek.81} W. H. Zurek, Pointer basis of quantum apparatus:  Into what 
      mixture does the wave packet collapse? \source{Phys. Rev. D}{24}{1516}{1981}.
\bibitem{Zurek.82} W. H. Zurek develops these points in the followup paper, 
       Environment-induced superselection rules, \source{Phys. Rev. D}{26}{1862}
        {1982}.
\bibitem{offdiagonal} It is characteristic of macroscopic systems that off-diagonal 
        elements of the reduced density matrix are not identically zero, but 
        unobservably small.
\bibitem{OCT} For a review see A. Bassi and G. C. Ghirardi, ``Dynamical reduction 
       models,'' \source{Phys. Reports}{379}{257}{2003}.
\bibitem{Cabello} A. Cabello,  Interpretations of quantum theory: A map of
       madness, eprint arXiv:1509.04711v2 [quant-ph] (2016).  See the list of
       ``Type II'' interpretations.
\bibitem{Esfeld} M. Esfeld, Essay review: Wigner's view of physical reality,
      \source{Stud. Hist. Phil. Mod. Phys.}{30B}{145}{1999}.
\bibitem{Bohr}  N. Bohr, Quantum mechanics and philosophy: causality and 
      complementarity, in {\it Philosophy in the Mid-Century}, (Ed.) R. Klibansky, 
      (La Nuova Italia Editrice, Florence, 1958). 
\bibitem{Camilleri} K. Camilleri, A history of entanglement: decoherence and
       the interpretation problem, \source{Stud. Hist. Phil. Mod. Phys.}{40}{290}  
       {2009}.
\bibitem{embrace} See the first paragraph on p. 74.\cite{Zeh}
\bibitem{ultimate}  From first three lines on p. 1523 of Zurek's 1981
       article.\cite{Zurek.81}
\bibitem{Weinbergpages} See pp. 86 - 95 of Weinberg's book.\cite{Weinbergbook}
\bibitem{Schloss} M. Schlosshauer, Decoherence, the measurement problem, and
       interpretations of quantum mechanics, \source{Rev. Mod. Phys.}{76}{1267}{2005}.
\bibitem{Joos} E. Joos and H. D. Zeh, The emergence of classical properties 
       through interaction with the environment, \source{Zeit. Phys. B: Condensed  
       Matter}{59}{223}{1985}.


\end{thebibliography}
\end{document}